\documentclass[fleqn,10pt]{wlscirep}
\usepackage[utf8]{inputenc}
\usepackage[T1]{fontenc}

\title{Quantum ghost imaging of a transparent polarisation sensitive phase pattern}
\author[1]{Aditya Saxena}\author[1]{Manpreet Kaur}\author[1]{Vipin Devrari}
\author[1*]{Mandip Singh}
\affil[1]{Department of Physical Sciences, Indian Institute of Science Education and Research (IISER) Mohali, Sector-81, Mohali, 140306, India.}

\affil[*]{mandip@iisermohali.ac.in}

\begin{abstract}
A transparent polarisation sensitive phase pattern exhibits a position and polarisation dependent phase shift of transmitted light and it represents a unitary transformation. A quantum ghost image of this pattern is produced with hyper-entangled photons consisting of Einstein-Podolsky-Rosen (EPR) and polarisation entanglement. In quantum ghost imaging, a single photon interacts with the pattern and is detected by a stationary detector and a non-interacting photon is imaged on a coincidence camera. EPR entanglement manifests spatial correlations between an object plane and a ghost image plane, whereas a polarisation dependent phase shift exhibited by the pattern is detected with polarisation entanglement. In this quantum ghost imaging, the which-position-polarisation information of a photon interacting with the pattern is not present in the experiment. A quantum ghost image is constructed by measuring correlations of the polarisation-momentum of an interacting photon  with polarisation-position of a non-interacting photon. The  experiment is performed with a coincidence single photon detection camera, where a non-interacting photon travels a long optical path length of 17.83~$m$ from source to camera and a pattern is positioned at an optical distance of 19.16~$m$ from the camera.

\end{abstract}
\begin{document}

\flushbottom
\maketitle
%
%
\thispagestyle{empty}

\section{Introduction}
Quantum ghost imaging of light absorbing objects relies on Einstein-Podolsky-Rosen (EPR) quantum entanglement of photons \cite{epr, gimage1, gimage6, shih_rev, shapiro}. A single photon after interaction with an absorptive object is detected by a stationary detector and a non-interacting photon is detected in the image plane of an imaging lens in a coincidence measurement with the interacting photon. EPR entanglement of photons manifests position correlations between the object plane and its ghost image plane. In quantum ghost imaging,  a momentum state of the interacting photon is measured in the object plane by placing a lens after the object plane, which focus the photon on a stationary single photon detector. Therefore, a photon detection does not reveal any information of location on the object where a photon has interacted with it. In other words, momentum measurement projects an interacting photon quantum state in a quantum superposition of its position states in the object plane. In contrast, a non-interacting photon is measured in a position basis in the image plane. Quantum ghost imaging techniques of photon-absorptive objects has been extensively investigated theoretically as well as experimentally \cite{gimage1,zeirev1, padgett_rev, shih_rev, shapiro, qimaging,eprboyd,duality, ghim, camera_pdgt, gi_timedomain,  gimage5, gimage6, malikboyd,int_gi}. Classical ghost imaging has been of considerable interest with thermal light \cite{shih, boyd, lugiato, gimage3}. Quantum imaging of spiral objects has been realised with random light \cite{chen1, yang1}. Furthermore, ghost imaging experiments are performed with X-rays \cite{xray1_gi, xray2_gi}, electrons \cite{elect_gi} and correlated atoms \cite{gitruscot1, gitruscot2}. A transparent phase object is imaged with quantum entangled photons using ghost diffraction \cite{imphase, p_sen_diff} and computational ghost imaging technique \cite{shapiro2, sdet_gi}. In a direct coincidence imaging experiment, a polarisation sensitive meta-surface is imaged with polarisation entangled photons \cite{br_pol}.

In this paper, a quantum ghost imaging experiment of a transparent polarisation sensitive phase pattern is presented, which manifests a polarisation and position dependent phase shift of transmitted photons. To image such type of pattern represented by a unitary transformation, quantum entanglement in the polarisation degree of freedom is incorporated in addition to EPR entanglement of photon pairs. Photons are EPR entangled and polarisation entangled separately in the form of a hyper-entangled state. Polarisation entanglement leads to detection of polarisation sensitive phase shift exhibited by the pattern and EPR entanglement correlates  position states of an interacting photon in the object plane with  position states of a non-interacting photon in the image plane. However, experiment presented in this paper is performed in quantum mechanical way, where the which-position-polarisation information of interacting photon on the pattern is not available even in principle. This experiment is a quantum ghost imaging version of a direct coincidence imaging experiment of a transparent polarisation sensitive phase pattern described in Ref. \cite{hype_m1}. However, in this paper the quantum entanglement under consideration involves EPR and polarisation degrees of freedom in hyper-entanglement of photon pairs, where in Ref.\cite{hype_m1} the hyper-entangled state involves momentum and polarisation degrees of freedom. These two experiments are fundamentally different. To produce a quantum ghost image, EPR-polarisation hyper-entanglement becomes necessary as further described in this paper. Therefore, experimental and theoretical analysis given in Ref. \cite{hype_m1} cannot be applied directly to the present experiment. Furthermore, the experiments described in this paper and in Ref.\cite{hype_m1} are very different from the experiment described in Ref.\cite{br_pol} which involves only polarisation entanglement of photon pairs and a direct imaging is performed in the close vicinity of object. In contrast, experiment presented in this paper and in Ref.\cite{hype_m1} are realised for a large optical path length, where the object is situated at a large distance from the imaging camera . Hyper-entanglement is necessary to realise a ghost and direct imaging of the distant object by jointly measuring momentum-polarisation of interacting photon and position-polarisation of non-interacting photon.

 As a consequence of EPR-polarisation hyper-entanglement the  polarisation, momentum and position quantum states of an individual photon are completely undefined prior to a measurement. Hyper-entangled photons are produced by type-II spontaneous parametric down conversion (SPDC) at the intersection of emission cones of down converted photons. A  nonlinear crystal is pumped by a broad diameter laser beam to produce the required quantum entangled state of photons. A polarisation sensitive phase pattern is produced by a spatial light modulator (SLM). A corresponding quantum ghost image is captured on an intensified-charge-coupled-device (ICCD) single photon sensitive camera.  ICCD camera detects a non-interacting photon only if an interacting photon is detected with a particular measurement outcome by a stationary single photon detector, which triggers the camera. A quantum ghost image gradually emerges after accumulating the detections of non-interacting photons by repeating the experiment for the same measurement outcomes. Each detection event of the non-interacting photon on the camera corresponds to its position measurement. However, momentum measurement outcome of interacting photon is kept same in all measurements presented in this paper, which reveals no position-information of the interacting photon on the pattern. Different quantum ghost images are produced for different polarisation detection outcomes of photons in a selected diagonal polarisation basis. By ignoring the interacting photon measurement outcomes or by letting this photon escape without any measurement, no ghost image is produced by single photon accumulation on camera. This is regarded as one of the main characteristics of quantum ghost imaging method, where an individual photon reveals no information of the pattern because of quantum entanglement.

\section{Concept and experiment}
\begin{figure*}
\begin{center}
\includegraphics[scale=1.38]{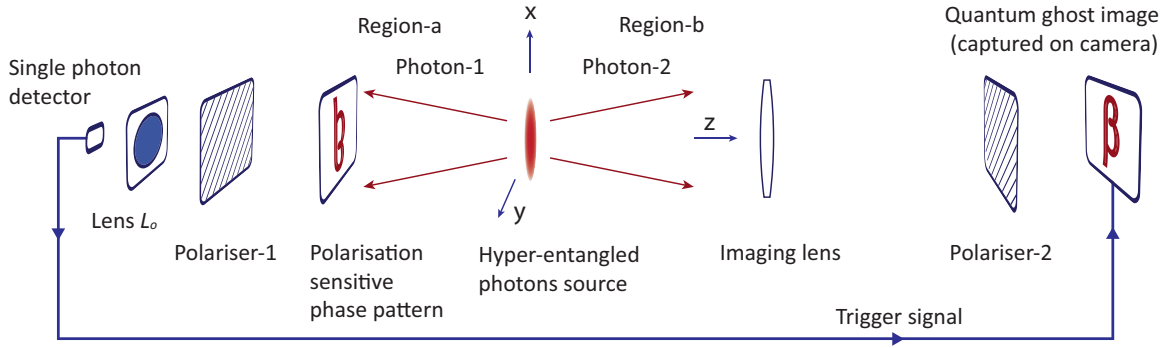}
\caption{\label{fig1} \emph{A schematic of a quantum ghost imaging experiment of a polarisation sensitive phase pattern. Photon-$1$ interacts with the pattern and detected by a single photon detector after passing through a polariser and a convex lens. A non-interacting photon-$2$ is imaged on camera after passing through an imaging convex lens and a polariser. Camera detects photon-$2$ leading to a joint measurement of its polarisation and position only if a particular measurement outcome of polarisation and momentum of photon-$1$ is produced.}}
\end{center}
\end{figure*}
Consider a schematic diagram of a ghost imaging configuration as shown in Fig.~\ref{fig1}, where a hyper-entangled photons pair is produced by a source positioned around origin. Photon-1 passes through a polarisation sensitive phase pattern without absorption. After that, it passes through a polariser and a convex lens $L_{o}$, which focus photon-$1$ on a single photon detector. If the detector detects  photon-1, then a particular joint measurement outcome corresponding to a specific polarisation and  momentum state of photon-$1$ in an object plane containing the pattern is produced. Whereas photon-2 never interacts with the pattern. It is detected  by a single photon sensitive coincidence camera, which measures its position in an image plane of an imaging convex lens after a polarisation selection made by a polariser-$2$.  Coincidence camera detects photon-2 only if it receives a trigger signal from a single photon detector \emph{i.e.} a signal corresponding to a particular joint measurement outcome of photon-1. An imaging lens is placed in path of photon-$2$ such that an object plane corresponds to a plane containing the pattern and an image plane corresponds to a plane of camera image sensor. Since the pattern under consideration is transparent and it introduces a polarisation dependent phase shift therefore, it cannot be imaged with conventional ghost imaging method for absorptive objects. The current setup shown in Fig.~\ref{fig1} differs from a conventional setup in three main aspects, (a) photons are polarisation entangled and EPR entangled separately, which is a necessary condition to produce a quantum ghost image of a transparent polarisation sensitive phase pattern. Polarisation entanglement detects the polarisation dependent phase shift and EPR entanglement correlates  locations of photons between the object and image planes. (b) The polarisation and  momentum state of photon-1 are measured  together in the object plane. (c) The polarisation and position states of photon-$2$ are measured together in the image plane. A quantum ghost image is produced by correlating the joint measurement outcomes of both photons.

In one-dimension,  EPR quantum state in position basis can be written as $|\alpha\rangle=\int^{\infty}_{-\infty}|x\rangle_{1}|x+x_{o}\rangle_{2} \mathrm{d}x$, where subscripts $1$ and $2$ are particle labels and $x_{o}$ is  position difference between position states of particles. In momentum basis, the same quantum state can be written as $|\alpha\rangle=\int^{\infty}_{-\infty}e^{i \frac{p x_{o}}{\hslash}}|p\rangle_{1}|-p\rangle_{2}\mathrm{d}p$, where particles have opposite momenta and $\hslash=h/2\pi$ is the reduced Planck's constant. Therefore, both position and momentum of each particle are completely unknown. If $x_{o}=0$, then both particles have same position in the integral. In addition, for the EPR quantum state $|\alpha\rangle$, both particles are equally likely to be found at all spatial and momentum locations \emph{i.e.}  spatial and momentum extensions of their position and momentum wavefunctions are infinite. In this paper, a three dimensional time independent scenario for photons is considered, where a finite source of photons is positioned around origin as shown in Fig.~\ref{fig1}. Let $\mathbf{r}'$ be an arbitrary point in the source where both photons are produced together. An amplitude to find photon-$1$ at  point $o_{a}$ in region-$a$ and photon-2 at point  $o_{b}$ in region-$b$ originating from a point $\mathbf{r}'$ and prior to any optical element can be written as $\frac{e^{ip_{1}|\mathbf{r}_{a}-\mathbf{r}'|/\hslash}}{|\mathbf{r}_{a}-\mathbf{r}'|}\frac{e^{ip_{2}|\mathbf{r}_{b}-\mathbf{r}'|/\hslash}}{|\mathbf{r}_{b}-\mathbf{r}'|}$ \cite{horne1, horne2}, where $p_{1}$ and $p_{2}$ are magnitudes of momentum of photon-$1$ and of photon-$2$, respectively. Consider $\mathbf{r}_{a}$ and $\mathbf{r}_{b}$ are position vectors
 of points $o_{a}$ and $o_{b}$ from origin, respectively. Therefore, $|\mathbf{r}_{a}-\mathbf{r}'|$ and $|\mathbf{r}_{b}-\mathbf{r}'|$ are distances of $o_{a}$ and $o_{b}$ from $\mathbf{r}'$, respectively.
In addition, $|H\rangle_{a}$ and $|V\rangle_{b}$ are the polarisation quantum states of photon-$1$ propagating in region-$a$ and of photon-$2$ propagating in region-$b$, respectively.

For an extended source, a quantum state leading to a finite amplitude to find a photon at $o_{a}$ with polarisation state $|H\rangle_{a}$ and a photon at $o_{b}$ with polarisation state $|V\rangle_{b}$ is a linear quantum superposition of amplitudes originating from all the points in the source, which can be written as
\begin{equation}
  \label{eq1}
   |\Psi\rangle_{12}=A_{o} \int^{\infty}_{-\infty}\int^{\infty}_{-\infty}\int^{\infty}_{-\infty} \psi(x',y',z')
  \frac{e^{ip_{1}|\mathbf{r}_{a}-\mathbf{r}'|/\hslash}}{|\mathbf{r}_{a}-\mathbf{r}'|}\frac{e^{ip_{2}|\mathbf{r}_{b}-\mathbf{r}'|/\hslash}}{|\mathbf{r}_{b}-\mathbf{r}'|} \mathrm{d}x'\mathrm{d}y'\mathrm{d}z' \otimes|H\rangle_a|V\rangle_b
\end{equation}

 where, $A_{o}$ is a constant and  $\psi(x',y',z')$ is an amplitude of pair creation at a location $r'(x',y',z')$. For a special case of an infinitely extended EPR state the amplitude of pair production $\psi(x',y',z')$ is constant for any arbitrary position vector $\mathbf{r}'$. A symbol $\otimes$ denotes separation of quantum states of different uncoupled degrees of freedom.
 Consider magnitudes of photons momenta are equal, $p_{1}=p_{2}=p$. Therefore, due to Bosonic nature of identical photons, a symmetric quantum state of photons is written as $|\Psi\rangle= \frac{1}{\sqrt{2}}(|\Psi\rangle_{12}+|\Psi\rangle_{21})$, where $|\Psi\rangle_{21}$ is a quantum state given in Eq.~\ref{eq1} after the exchange of photons.
Therefore, after symmetrization \cite{hype_m1}, a total symmetric quantum state can be written as
\begin{equation}
\label{eq2}
   |\Psi\rangle= A_{o} \int^{\infty}_{-\infty}\int^{\infty}_{-\infty}\int^{\infty}_{-\infty} \psi(x',y',z')
  \frac{e^{ip|\mathbf{r}_{a}-\mathbf{r}'|/\hslash}}{|\mathbf{r}_{a}-\mathbf{r}'|}\frac{e^{ip|\mathbf{r}_{b}-\mathbf{r}'|/\hslash}}{|\mathbf{r}_{b}-\mathbf{r}'|} \mathrm{d}x'\mathrm{d}y'\mathrm{d}z' \otimes\frac{1}{\sqrt{2}}(|H\rangle_1|V\rangle_2+|V\rangle_1|H\rangle_2)
\end{equation}

This integral represents a continuous variable entangled quantum state for a finite extension of EPR source and $\frac{1}{\sqrt{2}} (|H\rangle_1|V\rangle_2+|V\rangle_1|H\rangle_2)$ represents a discrete variable symmetric polarisation entangled state. Total quantum state in Eq.~\ref{eq2} is a hyper-entangled state, this combined quantum state is a tensor product of quantum entangled states in different degrees of freedom.

A pair creation amplitude is considered be a gaussian function such that $\psi(x',y',z')= a e^{-(x'^2+y'^2)/\sigma^2}e^{-z'^2/w^2}$, where $a$ is a constant, $\sigma$ and $w$ are the widths of the gaussian. Consider two planes parallel to $xy$-plane located on $z$-axis at distance $s_{1}$ (plane-$1$) and $s_{2}$ (plane-$2$) from origin. An amplitude to find photon-$1$ on plane-$1$ and photon-$2$ on plane-$2$ can be evaluated close to $z$-axis as follows. Consider distances of planes are such that $\sigma^{2}p/h s_{1}\geq 1$ and $\sigma^{2}p/h s_{2} \geq 1$, where $s_{1}$ and $s_{2}$ are much greater than $\sigma$ and $w$. This approximation is applicable to the experimental considerations presented in this paper. However, in Ref.\cite{hype_m1}, a small source extension is considered \emph{i.e.} $\sigma^{2}p/h s_{1}\ll1$ and $\sigma^{2}p/h s_{2} \ll 1$, which leads to a  momentum-polarisation hyper-entanglement of photons.

Therefore, Eq.~\ref{eq2} becomes

\begin{multline}
\label{eq3}
   |\Psi\rangle= \frac{a A_{o}}{s_{1}s_{2}} e^{i\frac{p(s_{1}+s_{2})}{\hslash}} \int^{\infty}_{-\infty}e^{-\frac{x'^2}{\sigma^2}} e^{\frac{ip}{\hslash}\left(\frac{(x_{1}-x')^2}{2s_{1}}+\frac{(x_{2}-x')^2}{2s_{2}}\right)} \mathrm{d}x' \int^{\infty}_{-\infty}e^{-\frac{y'^2}{\sigma^2}} e^{\frac{ip}{\hslash}\left(\frac{(y_{1}-y')^2}{2s_{1}}+\frac{(y_{2}-y')^2}{2s_{2}}\right)} \mathrm{d}y' \int^{\infty}_{-\infty}e^{-\frac{z'^2}{w^2}} e^{-i\frac{2p}{\hslash}z'}\mathrm{d}z'\\
  \otimes\frac{1}{\sqrt{2}}(|H\rangle_1|V\rangle_2+|V\rangle_1|H\rangle_2)
\end{multline}

After solving this integral,  Eq.~\ref{eq3} can be succinctly written as $|\Psi\rangle= \Phi(x_{1},y_{1}; x_{2}, y_{2}) \otimes (|H\rangle_1|V\rangle_2+|V\rangle_1|H\rangle_2)/\sqrt{2}$,
where $\Phi(x_{1},y_{1}; x_{2}, y_{2})$ is a two-photon position amplitude with variables in the argument separated by a semicolon denoting a position of photon-$1$ on plane-$1$ and of photon-$2$ on plane-$2$ such that

\begin{multline}
\label{eq5}
 \Phi(x_{1},y_{1}; x_{2}, y_{2})= c_{n}\frac{e^{i\frac{p}{\hslash}(s_{1}+s_{2})} e^{-i\phi}}{(4s^2_{1}s^{2}_{2}+(\frac{p}{\hslash})^{2}\sigma^{4} (s_{1}+s_{2})^{2})^{1/2}}
   \exp\left(\frac{-(\frac{p}{\hslash})^{2}(s_{1} s_{2} \sigma)^{2}\left((\frac{x_{1}}{s_{1}}+\frac{x_{2}}{s_{2}})^{2}+ (\frac{y_{1}}{s_{1}}+\frac{y_{2}}{s_{2}})^{2}\right)}{4s^2_{1}s^{2}_{2}+(\frac{p}{\hslash})^{2}\sigma^{4} (s_{1}+s_{2})^{2}}-\left(\frac{p}{\hslash}\right)^{2}w^{2}\right)\\
   \exp\left( \frac{-\frac{ip}{2 \hslash} \left((\frac{x_{1}}{s_{1}}+ \frac{x_{2}}{s_{2}})^{2} +  (\frac{y_{1}}{s_{1}}+ \frac{y_{2}}{s_{2}})^{2}\right)  (\frac{p}{\hslash})^{2}\sigma^{4} s_{1} s_{2} (s_{1}+s_{2}) }{4s^2_{1}s^{2}_{2}+(\frac{p}{\hslash})^{2}\sigma^{4} (s_{1}+s_{2})^{2}} \right)
   \exp\left(\frac{ip}{2 \hslash} \left(\frac{x^2_{1}+y^2_{1}}{s_{1}}+\frac{x^2_{2}+y^2_{2}}{s_{2}}\right)\right)
 \end{multline}

provided $\tan(\phi)=-p(s_{1}+s_{2})\sigma^{2}/2\hslash s_{1} s_{2}$ with $c_{n}$ as a constant.  The resulting quantum state of photons is a product of a continuous variable EPR entangled quantum state, where position states of photons on two planes are quantum entangled with each other, and a discretely entangled quantum state in the polarisation degree of freedom.  In this analysis to evaluate $\Phi(x_{1},y_{1}; x_{2}, y_{2})$, photons propagation close to $z$-axis is considered. However, by considering propagation along all directions in three-dimensional space for a large distance from the source, it is shown in Ref.\cite{hype_m1} that photons propagate in opposite direction along $z$-axis. Therefore, equation of plane-$1$ is $z=-s_{1}$ where photon-$1$ is found and of plane-$2$ is $z=s_{2}$ where photon-$2$ is found.
In actual experiment, anti-symmetric polarisation entangled state is produced as it remains same in the diagonal polarisation basis.  Photons are propagating in the opposite direction along $z$-axis from the source, in non-overlapping regions. This imposes a condition that $\Phi(x_{2},y_{2}; x_{1}, y_{1})=0$, when photons are further exchanged \cite{hype_m1}. Therefore, a hyper-entangled state produced by source can be written as
\begin{equation}
\label{eq5}
   |\Psi\rangle= \Phi(x_{1},y_{1}, z_{1}=-s_{1}; x_{2}, y_{2}, z_{2}=s_{2}) \otimes\frac{1}{\sqrt{2}}(|H\rangle_1|V\rangle_2-|V\rangle_1|H\rangle_2)
\end{equation}
where $z$-position of photons is incorporated in $\Phi(x_{1},y_{1}, z_{1}=-s_{1}; x_{2}, y_{2}, z_{2}=s_{2})$.

\subsection{Quantum ghost interference}  EPR entanglement part of Eq.~\ref{eq5} produces a quantum ghost interference \cite{ghost1, ghost2, ghost3}, which can be captured on a coincidence camera if a polarisation sensitive phase pattern is replaced with a double-slit positioned in a plane $z=-s_{1}$, which is plane-$1$. Since a double-slit under consideration is polarisation independent therefore, polarisation entanglement does not contribute to ghost interference.  Polariser-$1$ and polariser-$2$ can be removed to make no polarisation measurement. A convex lens $L_{o}$ focus photon-$1$ on a single photon detector. Detection of photon-$1$ at the detector projects quantum state of photon-$1$ onto a momentum state in a plane $z=-s_{1}$. In other words, this measurement projects photon-$1$ quantum state onto a quantum superposition of position states in a plane of the double-slit. This measurement does not reveal any which-path information at the double-slit. Interference pattern is produced in coincidence measurements of photons, where position of photon-$2$ is measured in a plane defined as an interference plane, which is same as  $z=s_{2}$ plane-$2$, located in path of photon-$2$. To obtain a quantum ghost interference, an imaging lens is removed and a camera detector surface is placed in the interference plane to detect position of photon-$2$. In another configuration, an imaging lens is placed such that its object plane corresponds to an interference plane and a corresponding image plane is further imaged by a telescope onto the camera. The latter configuration is preferable in experiments involving a long optical path.
\begin{figure}
\begin{center}
\includegraphics[scale=0.35]{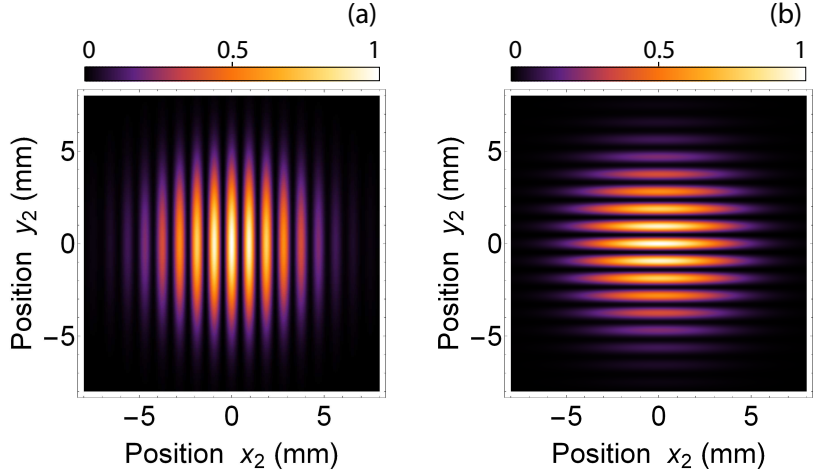}
\caption{\label{fig2} \emph{A quantum ghost interference pattern for a double-slit of slit separation $d=2$~mm. Double-slit is positioned such that (a) slit separation is along $x$-axis and (b) slit separation is along  $y$-axis.}}
\end{center}
\end{figure}
Consider a double-slit of slit separation $d$ along $x$-axis at $y=0$. A transmission function of the double-slit can be written as $A_{T}(x_{1},y_{1})= (\delta(x_{1}-d/2)\delta(y_{1})+\delta(x_{1}+d/2)\delta(y_{1}))/\sqrt{2}$. Therefore, an amplitude of coincidence detection of photon-$1$ by the detector and photon-$2$ in the interference plane at a location $(x_{2}, y_{2})$ can be written as
\begin{equation}
\label{eq6}
  A_{12}(x_{2}, y_{2})= \int^{\infty}_{-\infty}\int^{\infty}_{-\infty}A_{T}(x_{1}, y_{1})
   \Phi(x_{1},y_{1}, -s_{1}; x_{2}, y_{2}, s_{2}) \mathrm{d}x_{1} \mathrm{d}y_{1}
\end{equation}
Here, integration represents a projection onto a quantum superposition of position states of photon-$1$ in plane-$1$ containing a double-slit. Therefore, the probability of coincidence detection $P_{12}(x_{2}, y_{2})=|A_{12}(x_{2},y_{2})|^{2}$ represents a quantum ghost interference. Quantum ghost interference patterns shown in  Fig.~\ref{fig2} are obtained by solving Eq.~\ref{eq6}, for a double-slit located  at $z=-s_{1}=-1.33~m$ and an interference plane located at $z=s_{2}=1~m$. A slit separation of double-slit is $d=2~mm$ and $\sigma=3~mm$. In Fig.~\ref{fig2} (a), single slits of the double-slit are positioned at $x_{1}=1~mm$ and $x_{1}=-1~mm$ and in (b) single slits are positioned at $y_{1}=1~mm$ and $y_{1}=-1~mm$. A quantum ghost interference pattern originates due to EPR-entanglement of $\Phi(x_{1},y_{1}, -s_{1}; x_{2}, y_{2}, s_{2})$.

\subsection{Quantum ghost imaging of a polarisation sensitive phase pattern: Theory}
To  produce a quantum ghost image of a transparent polarisation sensitive phase pattern an imaging lens of focal length $f$ is positioned such that the pattern is located in its object plane $z=-s_{1}$ at a distance $u$ from  imaging lens, where $u>f$. Coincidence camera is positioned in an image plane of imaging lens located at a distance $v$ from the lens. Imaging lens and pattern are positioned on opposite locations \emph{w.r.t.} source as shown in Fig.~\ref{fig1}.  Therefore, position of imaging lens from origin is $s_{2}=u-s_{1}$. Neither of the photons reveal information of the pattern therefore, a quantum ghost image is obtained by making correlated joint measurements on both photons.

Consider, a cartesian coordinate $(\xi,\eta)$-plane, coinciding with a plane $z=s_{2}$ with origin on $z$-axis, where a thin imaging lens is positioned such that $\xi$-axis is parallel to $x$-axis and $\eta$-axis is parallel to $y$-axis. A polarisation sensitive phase pattern transforms the polarisation quantum state of photon-$1$ as a unitary transformation \emph{i.e.} $|H\rangle_{1}\mapsto e^{i\phi(x_{1}, y_{1})}|H\rangle_{1}$ and $|V\rangle_{1}\mapsto|V\rangle_{1}$ at a position $(x_{1}, y_{1})$ on the pattern. Where $\phi(x_{1}, y_{1})$ is an arbitrary position dependent phase shift introduced by the pattern. This transformation couples the polarisation entanglement with a position of photon-$1$ on the pattern. A quantum image state is defined as a quantum state that results in coincidence detection of photon-$1$ at a position $(x_{1}, y_{1})$ just after passing through the pattern and photon-$2$ at a position ($x_{2}, y_{2}$) in the image plane of imaging lens but without polarisation measurement.  A quantum image state can be written as

\begin{multline}
\label{eq7}
  |\Psi\rangle_{I}=c_{I} \int^{\infty}_{-\infty}\int^{\infty}_{-\infty}\Phi(x_{1},y_{1}, -s_{1}; \xi,\eta, s_{2}) a_{p2}(\xi,\eta) e^{-i\frac{p}{\hslash}\frac{(\xi^{2}+\eta^{2})}{2 f}}
    e^{i\frac{p}{\hslash}\left(\frac{(x_{2}-\xi)^{2}}{2v}+\frac{(y_{2}-\eta)^{2}}{2 v}\right)}\mathrm{d}\xi\mathrm{d}\eta\\
    \frac{1}{\sqrt{2}}(e^{i\phi(x_{1},y_{1})}|H\rangle_1|V\rangle_2-|V\rangle_1|H\rangle_2)
\end{multline}
which is succinctly written as $|\Psi\rangle_{I}= \Phi_{I}(x_{1}, y_{1}; x_{2}, y_{2}) (e^{i\phi(x_{1},y_{1})}|H\rangle_{1}|V\rangle_{2}-|V\rangle_{1}|H\rangle_{2})/\sqrt{2}$, where $c_{I}$ is a constant, $a_{p2}(\xi,\eta)$ is a transmission function of an aperture placed close to the imaging lens, $e^{-ip(\xi^{2}+\eta^{2})/2f\hslash}$ is the phase shift term introduced by the imaging convex lens. After transmission through the lens, photon-$2$ propagates to the imaging plane where propagation from a position ($\xi,\eta$) to a position ($x_{2}, y_{2}$) is given by the term
$e^{ip((x_{2}-\xi)^{2}+(y_{2}-\eta)^{2})/2v\hslash}$. For a circular lens of radius $\rho$ without any additional aperture, the transmission aperture function $a_{p2}(\xi,\eta) = 1$ for $\xi^{2}+\eta^{2}\leq\rho^{2}$ and zero otherwise.

A quantum ghost image is obtained by correlating the joint measurements of polarisation and momentum of photon-$1$ in the object plane with measurements of polarisation and position of photon-$2$ in the image plane. Consider a diagonal polarisation basis for  photon-$j$ ($j=1$ or $2$) such that $|d^{+}\rangle_{j}=(|H\rangle_{j}+|V\rangle_{j})/\sqrt{2}$ and $|d^{-}\rangle_{j}=(|H\rangle_{j}-|V\rangle_{j})/\sqrt{2}$, where $|d^{+}\rangle_{j}$, $|d^{-}\rangle_{j}$ represent linear polarisation states of photon-$j$ with planes of polarisation at $\delta_{j}=+45^{o}$ and $\delta_{j}=-45^{o}$ \emph{w.r.t.} $x$-axis, respectively.
If pass axes of polariser-$1$ and polariser-$2$ are oriented at $\delta_{1}=\delta_{2}=-45^{o}$ then photon-$1$ is measured in a polarisation state  $|d^{-}\rangle_{1}$ and photon-$2$ is measured in a polarisation state  $|d^{-}\rangle_{2}$ after their detection. Therefore, from Eq.~\ref{eq7} the probability of coincidence detection corresponding to a polarisation measurement outcome $(|d^{-}\rangle_{1}, |d^{-}\rangle_{2})$ can be calculated as

\begin{equation}
\label{eq8}
 P_{d^{-}_{1}, d^{-}_{2}}(x_{2},y_{2})=|\int^{\infty}_{-\infty}\int^{\infty}_{-\infty} {_{2}\langle d^{-}|} {_{1}\langle d^{-}|}\Psi\rangle_{I} a_{p1}(x_{1},y_{1})\mathrm{d}x_{1}\mathrm{d}y_{1}|^{2}
\end{equation}
which can be written as
\begin{equation}
\label{eq9}
 P_{d^{-}_{1}, d^{-}_{2}}(x_{2},y_{2})=|\int^{\infty}_{-\infty}\int^{\infty}_{-\infty} a_{p1}(x_{1},y_{1}) \Phi_{I}(x_{1}, y_{1}; x_{2}, y_{2})
 \left(\frac{1-e^{i\phi(x_{1}, y_{1})}}{2\sqrt{2}}\right)\mathrm{d}x_{1}\mathrm{d}y_{1} |^{2}
\end{equation}
where $a_{p1}(x_{1},y_{1})$ is the aperture transmission function of an aperture placed just after the pattern being imaged.

Similarly, if polarisers are aligned such that $\delta_{1}=+45^{o}$ and $\delta_{2}=-45^{o}$ for $(|d^{+}\rangle_{1}, |d^{-}\rangle_{2})$ measurement, then the probability of coincidence detection of photons can be written as
\begin{equation}
\label{eq10}
 P_{d^{+}_{1}, d^{-}_{2}}(x_{2}, y_{2})=|\int^{\infty}_{-\infty}\int^{\infty}_{-\infty} a_{p1}(x_{1},y_{1})\Phi_{I}(x_{1}, y_{1}; x_{2}, y_{2})
 \left(\frac{1+e^{i\phi(x_{1}, y_{1})}}{2\sqrt{2}}\right)\mathrm{d}x_{1}\mathrm{d}y_{1}|^{2}
\end{equation}
 The coincidence detection probability represents a quantum ghost image of a polarisation sensitive phase pattern. Quantum  ghost images $P_{d^{-}_{1}, d^{-}_{2}}(x_{2},y_{2})$ and $P_{d^{+}_{1}, d^{-}_{2}}(x_{2},y_{2})$ are inverted \emph{w.r.t} each other.

\subsection{Experiment}
\begin{figure*}
\begin{center}
\includegraphics[scale=1.15]{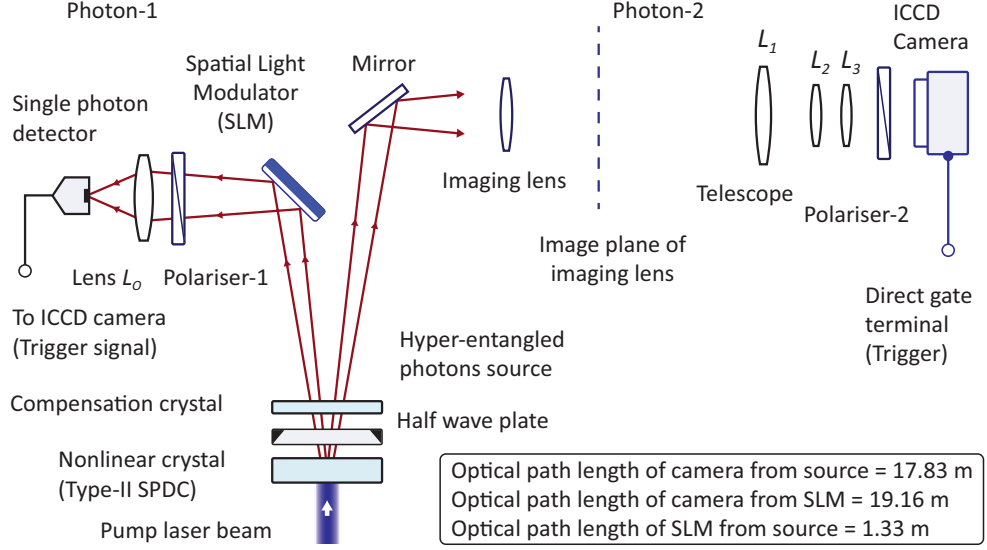}
\caption{\label{fig3} \emph{A schematic of experiment of quantum ghost imaging of a polarisation sensitive phase pattern. This is an experimental configuration of a schematic diagram shown in Fig.~\ref{fig1}. The diagram is not to the scale and photon-$1$ is incident on SLM close to a normal incidence.}}
\end{center}
\end{figure*}

\begin{figure*}
\begin{center}
\includegraphics[scale=0.3]{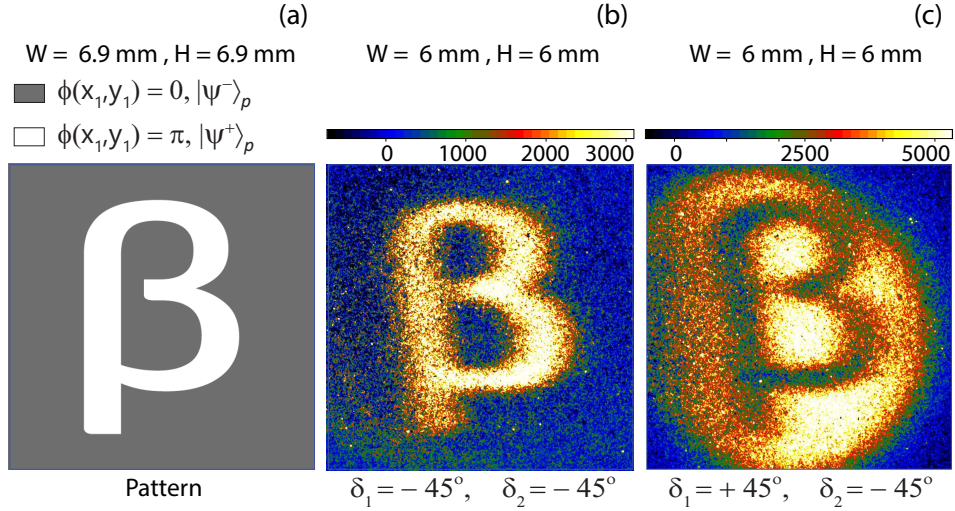}
\caption{\label{fig4} \emph{(a) A polarisation sensitive phase pattern. (b) A quantum ghost image of the pattern for $\delta_{1}=-45^{o}$ and $\delta_{2}=-45^{o}$. (b) An inverted quantum ghost image of the pattern for $\delta_{1}=+45^{o}$ and $\delta_{2}=-45^{o}$. Distance of pattern from camera is 19.16~m.}}
\end{center}
\end{figure*}
A polarisation sensitive phase pattern under consideration is shown in Fig.~\ref{fig4} (a), which is produced by SLM. A lighter region on the pattern introduces a phase shift $\phi(x_{1},y_{1})=\pi$ for horizontally polarised photon-$1$ and a darker region corresponds to $\phi(x_{1},y_{1})=0$. After a transformation by the pattern \emph{i.e.} $|H\rangle_{1}\mapsto e^{i\phi(x_{1}, y_{1})}|H\rangle_{1}$ and $|V\rangle_{1}\mapsto|V\rangle_{1}$, a polarisation entangled state of photons becomes $|\Psi^{-}\rangle_{p}=(|H\rangle_1|V\rangle_2-|V\rangle_1|H\rangle_2)/\sqrt{2}$ for $\phi(x_{1},y_{1})=0$ and $|\Psi^{+}\rangle_{p}=-(|H\rangle_1|V\rangle_2+|V\rangle_1|H\rangle_2)/\sqrt{2}$ for $\phi(x_{1},y_{1})=\pi$.  A schematic of experiment performed is shown in Fig.~\ref{fig3}. Hyper-entangled photons are produced by type-II SPDC process in a beta-barium-borate (BBO) nonlinear anisotropic crystal. A broad beam diameter pump laser light of wavelength $405~nm$ is incident on the crystal, which is aligned to produce non-collinear degenerate $810~nm$ photons. Photons are hyper-entangled at the intersection of conical emission profiles of down converted photons. Intersection regions expand with distance from BBO crystal and width of each region is about $7~mm$ in a transverse plane at a distance $1.33~m$ from the crystal. EPR entanglement and polarisation entanglement are produced at the intersection regions. A compensation crystal is placed after the first BBO crystal to compensate for longitudinal and transverse walk-offs.  Compensation crystal is aligned parallel to the first crystal such that optic axes of both crystals are parallel to each other and both photons pass through the compensation crystal. A half wave-plate is inserted between the two crystals to interchange orthogonal linear polarisation states. Photon-$1$ of an entangled pair is incident on SLM close to a normal incidence, which produces a polarisation sensitive phase pattern. Difference between angle of reflection and angle of incidence on SLM is about $8^{o}$. After passing through a convex lens $L_{o}$, polariser-$1$ and an optical bandpass filter of center wavelength $810~nm$, photon-$1$ is detected by a single photon detector. Photon-$2$ never interacts with the pattern and it passes through an imaging convex lens. Imaging lens is positioned such that it produces an image of SLM surface onto the image plane as shown in Fig.~\ref{fig3}. This image plane acts as an object for a telescope, which produces its image on ICCD camera. Photon-$2$ travels a longer optical distance as compared to photon-$1$ in order to provide time to ICCD camera to open its direct gate electronic shutter after receiving a trigger signal from single photon detector. ICCD camera direct gate shutter opening delay time is $20~ns$ after receiving a trigger signal. Optical distance of ICCD camera from BBO crystal is $17.83~m$ and an optical distance of SLM surface from ICCD camera is $19.16~m$. Optical distance of SLM surface from BBO crystal is $1.33~m$. Focal length of imaging lens is $1.5~m$, this lens images SLM surface onto the image plane as shown in Fig.~\ref{fig3} however, imaging lens distance from BBO crystal is same as its focal length to collimate non-interacting photons on the telescope, otherwise photon coincidence counts reduce considerably.

Telescope consists of three convex lens system, where lens $L_{1}$ of focal length $40~cm$ and diameter $5~cm$ produces a real image of an image plane of the imaging lens. Lens $L_{2}$ of focal length $10~cm$ produces a virtual image of the real image produced by lens $L_{1}$. Lens $L_{3}$ of focal length $f_{3}= 1~m$ is configured in a $2f_{3}$ imaging configuration such that ICCD camera detector surface is placed at a distance $2~m$ from lens $L_{3}$ to image the virtual image produced by lens $L_{2}$. Since a virtual image is displaced by displacing lens $L_{2}$ therefore, a sharp image is produced on ICCD camera sensor surface if virtual image is located at a distance $2f_{3}$ from lens $L_{3}$. In this configuration, only lens $L_{2}$ is displaced to image any plane under consideration. Polariser-$2$ is placed close to ICCD camera after the telescope. Polariser-$1$ is placed at a distance $22~cm$ from SLM as shown in Fig.~\ref{fig3}.

Polarisation entanglement part of the hyper-entangled state is characterised by violating Clauser, Horne, Shimony, Holt (CHSH) inequality \emph{i.e.} $0<|S|<2$ \cite{chsh_th, chsh, hype_m1}. Experimentally measured value of parameter is $S=-2.57$. A polarisation sensitive phase pattern is shown in Fig.~\ref{fig4} (a). A quantum ghost image of the pattern is constructed by measuring coincidence photon counts between single photon detector and different locations on ICCD camera. Prior to any quantum ghost image measurement, a coincidence background ghost image is captured on ICCD camera by imprinting a uniform polarisation sensitive phase pattern on SLM that leads to minimum coincidence photon counts for the same inclination of polarisers that is $\delta_{1}=\delta_{2}=-45^{o}$. This background image is subtracted from all subsequent images to obtain a quantum ghost image. A polarisation sensitive phase pattern is shown in Fig.~\ref{fig4}(a), where a lighter region transforms the polarisation entangled state to $|\Psi^{+}\rangle_{p}$ and a darker region represents an identity transformation resulting same polarisation entangled state $|\Psi^{-}\rangle_{p}$ as produced by the source. A quantum ghost image of this polarisation sensitive phase pattern is shown in Fig.~\ref{fig4} (b) for same inclination of pass axes of polarisers \emph{i.e.} $\delta_{1}=\delta_{2}=-45^{o}$.  The quantum ghost image of the same pattern gets inverted as shown in Fig.~\ref{fig4} (c), when pass axis of polariser-$1$ is inclined at $\delta_{1}=+45^{o}$ and pass axis of polariser-$2$ remains at same inclination, $\delta_{2}=-45^{o}$. A coincidence background image and a coincidence image with the pattern under consideration are taken for $30~min$ time of exposure of ICCD camera under the direct gate operation in accumulation mode. In accumulation mode,  frames captured during the exposure time are accumulated. However, a frame can register photon-2 only if direct gate shutter is opened by the trigger signal during its acquisition. Negative coincidence counts regions in Fig.~\ref{fig4} (b) and (c) are due to experimental fluctuations because each quantum ghost image is constructed by subtracting a background image from an image with the pattern. The quantum ghost image shown in Fig.~\ref{fig4}(c) is an inverted image \emph{w.r.t.} the quantum ghost image shown in Fig.~\ref{fig4}(b).  Demagnification of each image is 0.87. Single photon-$2$ accumulation on ICCD camera without making any correlation with photon-$1$ measurements results in no image formation.

\section{Conclusion}
This paper presented an experiment on quantum ghost imaging of a transparent polarisation sensitive phase pattern with EPR-polarisation hyper-entangled photons. Since the pattern is polarisation sensitive therefore, polarisation entanglement is incorporated  to detect a polarisation sensitive phase exhibited by the pattern. A quantum ghost image is a combined result of EPR entanglement and polarisation entanglement of photons.  Paper presented a detailed analysis of hyper-entanglement of photons, quantum ghost interference and quantum ghost imaging. The required hyper-entangled state is produced with type-II SPDC process in a BBO crystal. The experiment is performed with a gated ICCD single photon sensitive camera to capture coincidence ghost images. Each quantum ghost image is background corrected. Two joint measurements on each photon, which correspond to polarisation, momentum of photon-$1$ and polarisation, position of photon-$2$, are correlated to construct a quantum ghost image.

\section{Methods}
Hyper-entangled photon pairs are produced by Type-II degenerate SPDC process in a BBO crystal at 810~$nm$. This source configuration is different from the configuration used in a previously published quantum imaging experiment based on hyper-entangled photon pairs, where a hyper-entangled state consists of momentum entanglement and polarisation entanglement, which is produced at a very large distance from the source as described in Ref.\cite{hype_m1}. In the present paper, hyper-entangled state consists of EPR-entanglement and polarisation entanglement, which is produced at the intersection of down converted cones. However, in the present source the beam size of the pump laser beam is approximately expanded by twenty times as compared to the source described in Ref.\cite{hype_m1}. In this configuration, the integral given by Eq.~\ref{eq2} cannot be approximated as given in the previous paper Ref.\cite{hype_m1}.  EPR entanglement produces quantum ghost interference however, momentum entanglement cannot produce quantum ghost interference. EPR-entanglement and polarisation entanglement produce a quantum ghost image of the polarisation sensitive phase pattern. Typical pump laser power passing through the BBO crystal is 60~$mW$ at 405~$nm$. A walk-off compensation crystal of half thickness is placed parallel to the first crystal and a half wave plate is placed between the two crystals to interchange orthogonal linear polarisation states.   Diameter of each intersection region of cones expands with distance from the crystal. The required EPR-polarisation hyper-entangled state is produced at the intersection regions. No image is observed for non-interacting photon accumulation on ICCD camera without correlating it with the interacting photon. In a coincidence measurement, the direct gate electronic shutter of ICCD camera is opened by a trigger pulse generated by the single photon detector after detection of an interacting photon. Direct gate shutter opening allows a detection of photon by the ICCD camera. Direct gate shutter remains open for the duration of trigger pulse which is $10~nsec$. Single photon detector produces $2\times10^{4}$ trigger pulses per second. Therefore, the direct gate shutter opens $3.6\times10^{7}$ times in 30~min while camera is accumulating frames continuously.  A narrow time window and low photon pair generation rate allow a selective detection of an entangled pair of photons \emph{i.e.} it significantly reduces the overlapping of detection events of a selected pair of photons with another pairs. Ghost images shown in Fig.~\ref{fig4} correspond to accumulation of photon counts in the coincidence window for $30~min$ time of frame-accumulation on the ICCD camera. A separate background ghost image of a uniform pattern is taken under same conditions. Background image is subtracted from each image of a pattern under consideration to obtain its background corrected ghost image shown in Fig.~\ref{fig4}. Image accumulation and background substraction are performed by the ICCD camera.

\section{Acknowledgments} {Mandip Singh acknowledges research funding by the Department of Science and Technology, Quantum Enabled
Science and Technology grant for project No. Q.101 of theme title “Quantum Information Technologies with
Photonic Devices”, DST/ICPS/QuST/Theme-1/2019 (General).}

\section{Author contributions}
 MS setup the experiment. MK, AS and MS performed the experiment. MK, AS and VD took data. MS developed concept and theory. MS drawn diagrams and wrote the manuscript.

\section{Competing interests}

The authors declare no competing interest.

\section{Data availability statement}

All data generated or analysed during this study are included in this published article.


\bibliography{sample}

\end{document}